\documentclass{article}
\usepackage[toc,page]{appendix}
\usepackage[authordate, backend=biber, uniquename=false, mincitenames=1, maxcitenames=1,citetracker,natbib]{biblatex-chicago}

\usepackage[a4paper, total={6in, 8in}]{geometry}
\usepackage[utf8]{inputenc}
\usepackage{amsmath}
\usepackage{amsthm,amssymb}
\DeclareMathOperator*{\argmin}{arg\,min}
\usepackage{booktabs} 
\usepackage{caption}
\usepackage{setspace}
\usepackage{graphicx}
\usepackage{comment}
\usepackage{lscape}
\usepackage{amsfonts}

\newtheorem{prop}{Proposition}
\newtheorem{definition}{Definition}
\usepackage{amsthm}
\usepackage{subcaption}
\usepackage[ruled,vlined]{algorithm2e}
\usepackage{booktabs,caption}
\usepackage[flushleft]{threeparttable}
\usepackage[labelfont=bf]{caption}
\usepackage{subcaption}
\usepackage{hyperref} 
\hypersetup{colorlinks,citecolor=blue,allcolors=blue}

\DeclareFieldFormat{citehyperref}{%
  \DeclareFieldAlias{bibhyperref}{noformat}
  \bibhyperref{#1}}

\DeclareFieldFormat{textcitehyperref}{%
  \DeclareFieldAlias{bibhyperref}{noformat}
  \bibhyperref{%
    #1%
    \ifbool{cbx:parens}
      {\bibcloseparen\global\boolfalse{cbx:parens}}
      {}}}

\savebibmacro{cite}
\savebibmacro{textcite}

\renewbibmacro*{cite}{%
  \printtext[citehyperref]{%
    \restorebibmacro{cite}%
    \usebibmacro{cite}}}

\renewbibmacro*{textcite}{%
  \ifboolexpr{
    ( not test {\iffieldundef{prenote}} and
      test {\ifnumequal{\value{citecount}}{1}} )
    or
    ( not test {\iffieldundef{postnote}} and
      test {\ifnumequal{\value{citecount}}{\value{citetotal}}} )
  }
    {\DeclareFieldAlias{textcitehyperref}{noformat}}
    {}%
  \printtext[textcitehyperref]{%
    \restorebibmacro{textcite}%
    \usebibmacro{textcite}}}

\title{Robust PCA Synthetic Control}
\author{Mani Bayani}
\date{August 2021}

\begin{document}
\maketitle
\begin{abstract}
\noindent In this study, I propose an algorithm for synthetic control method for comparative studies. My algorithm builds on the synthetic control model of \cite{abadie2015} and the later model of \cite{Amjad}. I apply all three methods (robust PCA synthetic control, synthetic control, and robust synthetic control) to answer the hypothetical question, what would have been the per capita GDP of West Germany if it had not reunified with East Germany in 1990? I then apply all three algorithms in two placebo studies. Finally, I check for robustness. This paper demonstrates that my method can outperform the robust synthetic control model of \cite{Amjad} in placebo studies and is less sensitive to the weights of synthetic members than the model of \cite{abadie2015}. \\
\end{abstract}
\section{Introduction}
Since first introduced by \cite{abadie2003}, synthetic control (SC) models have attracted a great deal of attention as a powerful method for comparative studies in the absence of observations for a treated unit. The basic idea behind synthetic control is to find a suitable synthetic group and estimate this group's relation with the treated unit for the pre-intervention period. Then, the synthetic group's weights can be used for predicting the treated unit's outcome as if it had never been treated. The practice of estimating the difference between the outcome of a policy on a treated unit and its counterfactual has been growing in many fields, such as public health (\cite{Bouttell}) and criminology (\cite{Goh}). As such, scholars over the last two decades have worked to improve this method. Among the most notable of these studies, \cite{Athey2017} use the matrix norm and matrix completion methods for estimating the counterfactual treated unit. In a study that is closely related to my method, \cite{Amjad} use the combination of de-noising technique and singular value thresholding to estimate the underlying linear relation between a treated unit and its synthetic before the treatment.\\ 
In line with \cite{Amjad}, I propose in this study a five-step, data-driven algorithm that can find the underlying linear relation between a treated unit and its synthetic. Given a data set $\mathcal{Y}$ of the outcome variable that includes both treated unit $\mathcal{Y}_i$ and untreated units $\mathcal{Y}_{-i}$, for the pre-intervention period $1,...,T_0$ and post intervention period $T_0+1,...,T$ the robust PCA synthetic control first computes the functional principle component scores of $\mathcal{Y}$ for the pre-intervention period, then applies the k-means algorithm over these scores. The untreated units that fall in the same cluster as the treated unit are considered the donor pool ($Y_{-i}$) of the treated unit ($Y_{i}$). This method then uses the robust PCA method to extract the low-rank structure of the donor pool ($L$). After that, a linear relation between the treated unit and the low-rank structure of the donor pool is computed by a simple optimization for the pre-intervention period $1,...,T_0$. This linear relation can then be used to estimate the counterfactual of the treated unit for the post intervention period $T_0 +1,...,T$. \\
The contribution of this study is in two folds. First, my method can automatically separate the donor pool from the non relevant units given any data set by using unsupervised learning. Second, to extract the low-rank structure of the donor pool, the algorithm uses robust PCA that is not sensitive to outliers and missing data.\\
This article is organized as follows. In the next section, I introduce the concept and theories of the first $3$ steps of robust PCA synthetic control: 1.) Functional principle component scores, 2.) K-means algorithm, and 3.) Robust principle component analysis. Then in section \ref{sec_2.4}, I explain the five-step algorithm of robust PCA synthetic control in detail. In section \ref{sec3}, I apply my method to the case of West Germany's reunification in $1990$ to find the impact of reunification on West Germany's per capita GDP. In addition to comparing the output of robust PCA synthetic with \cite{abadie2015} and robust synthetic control suggested in \cite{Amjad}, I check the placebo studies and robustness of all three models. In section \ref{sec4}, I run a simulation study to check the accuracy of my method. Finally, the section \ref{sec5} concludes the article.

\section{Methodology}
As previously mentioned, I use a five-step algorithm to estimate a linear relation between the donor pool and the treated unit for the pre-intervention time period, and then estimate a counterfactual of the treated unit for the post-intervention time period. In the first step, I adopt the functional principal components analysis (FPCA) for the pre-intervention time period to overcome the curse of dimensionality in data and prepare it for the K-means algorithm. FPCA finds a set of orthogonal bases that maximize the variance of the original data. Then, I project the original data on these orthogonal bases to find the FPCA scores. Choosing $k << T_0$ of these FPCA scores, we can reduce the dimension of our data from $T_0$ to $k$ and then use these low-dimension data for clustering with K-means algorithm. Data points that fall in the same cluster as the treated unit are considered as the donor pool and I use the robust principle component analysis to extract its low-rank structure. Next, this low-rank structure of the donor pool will be used for estimating the linear relation between the treated unit and its synthetic. Finally, this linear relation predicts the counterfactual of the treated unit for the post-intervention time period. I explain each of these steps separately in the following sections.

\subsection{Step 1: Computing Functional Principal Component}

The theory related to the functional data analysis (FDA) was introduced decades ago and has been in use since the 1960s (\cite{Shang2015}). With the recent development of efficient computing, FDA has become widespread in many fields \footnote{For more information about recent FDA applications, see \cite{Aullah}.}. One application of FDA is to find the low-rank structure of high-dimensional functional data, called functional principle components (FPC). The classic principle components can be computed by eigendecomposition of the covariance matrix $\Sigma$: 
\begin{align}
    \Sigma = U D U^T = \sum_{i=1}^r d_i u_i u_i^T
\end{align}
where $U$ is the matrix with eigenvectors of $\Sigma$ as its columns, $D$ is the diagonal matrix with eigenvalues of $\Sigma$ on its diagonal, $r$ is the rank of the covariance matrix and $U^TU=I$. But as discussed by \cite{croux} and \cite{Shang2015}, the classic principle component analysis can be misleading in the presence of panel data, sparsity, or when the dimension of data points is greater than the sample size. Counterfactual estimation may have all three of these issues: counterfactual estimation always uses panel data, the pre-intervention time period is often greater than the number of synthetic units (meaning the dimension of data points is greater than the sample size), and there may be missing values. To overcome these potential issues, I suggest using FPCA for dimensional reduction. \\
The basic idea of FPCA is similar to PCA, but instead of eigenvectors of the covariance matrix, FPCA uses the eigenfunctions of covariance function to reduce the dimensionality.
If we assume that data point $j$, $\mathcal{Y}^-_j(t)$\footnote{The negative superscript is for the pre-intervention period, so $\mathcal{Y}^-(t) \in \mathcal{R}^{M \times T_0}$ contains the outcome variable of treated unit and non treated units for the pre-intervention period.}, is a $L^2$ random process defined over close interval $[a,b]$, we can decompose $\mathcal{Y}^-_j(t)$ as:
\begin{align}
    \mathcal{Y}^-_j(t) = \mu(t) + \epsilon_j(t)
\end{align}
where $\mu(t)$ is the continuous mean and $\epsilon_j(t)$ is a realization from a stochastic process with the mean zero and the covariance function $\sigma(t,t')$. $\epsilon_j(t)$ includes both random noise and signal-to-signal variations. Based on the Karhunen-Loeve theorem, we can write the $\epsilon_j(t)$ as:
\begin{align}
    \epsilon_j(t) = \sum_{k=1}^{\infty} \xi_{jk} \phi_k(t)
\end{align}
where $\xi_{jk}$ are the zero mean, uncorrelated coefficients with finite variances and $\phi_k(t)$ are eigen-functions of the covariance function $\sigma(t,t') = cov(\epsilon(t),\epsilon(t'))$. Therefore, we have:
\begin{align}
\label{eigen}
    \sigma(t,t') = \sum_{k=1}^{\infty} \lambda_k \phi_k(t) \phi_k(t')
\end{align}
In equation \eqref{eigen}, $\lambda$'s are ordered eigenvalues and with this setting, eigenfunctions of covariance function can be obtained by:
\begin{align}
    \int_a^b \sigma(t,t') \phi_k(t) dt = \lambda_k \phi_k(t')
\end{align}
There are several methods to estimate the covariance and eigenfunction\footnote{For a list of possible methods for FPCA, see \cite{Aullah}.}. Here I follow the \cite{LiHsing} method that is robust to the missing or sparse data. If we assume that $\xi_{jk}$ decay fast with $k$, we only need a finite number of $\xi_{jk}$ to estimate the data point by:
\begin{align}
    \mathcal{Y}^-_j(t) = \mu(t) + \epsilon_j(t) \cong \mu(t) + \sum_{k=1}^{K} \lambda_k \phi_k(t) \phi_k(t')
\end{align}
Following \cite{LiHsing}, we can estimate the mean function $\hat \mu(t)$ by a local linear regression by the following minimization:
\begin{align}
\label{mean_fpca}
    \min_{\beta_0,\beta_1} \sum_{j=1}^M \sum_{k=1}^{T_0} K\left(\frac{t_{jk}-t}{h}\right) \left [\mathcal{Y}^-_j(t_{jk})-\beta_0 - (t - t_{jk})\beta_1 \right]^2
\end{align}
where $j=1,2,..,M$ is the unit index, $k=1,2,...,T_0$ is the time index in the each unit and $K(.)$ is a Kernel function. After we have the estimation of mean function by \eqref{mean_fpca}, we can use it to find the raw covariance function of $j$th unit, $\hat \sigma_j(t,t')$ by:
\begin{align}
  \hat \sigma_j(t_{jk},t_{js}) = (\mathcal{Y}^-_j(t_{jk})-\hat \mu(t_{jk})) (\mathcal{Y}^-_j(t_{js})-\hat \mu(t_{js})) \hspace{0.5 cm} j= 1,...,M \hspace{0.1 cm} ,k \neq s
\end{align}
Then, following \cite{LiHsing}, we can estimate the covariance surface $\hat \sigma(t,t')$ by a local quadratic regression according to the following minimization:
\begin{align}
\begin{split}
   \min_{\beta_0,\beta_1,\beta_2} \sum_{j=1}^M \sum_{1 \leq k \neq s \leq T_0} &K\left(\frac{t_{jk}-t}{h},\frac{t_{js}-t'}{h}\right). \\& \left [\hat \sigma_j(t_{jk},t_{js})-\beta_0 - (t - t_{jk})\beta_1 - \beta_2 (t' - t_{js})^2 \right]^2 
\end{split}
\end{align}
After computing the surface of covariance, we can compute the eigenfunction $\hat \phi_k(t_j)$ by solving the following equation:
\begin{align}
   \label{intergral}
    \int_a^b \hat \sigma(t,t') \hat \phi_k(t) dt =\hat \lambda_k \hat \phi_k(t')
\end{align}
where $\int_a^b \hat \phi_k(t) \times \hat \phi_m(t)$ is equal to one for $m=k$ and zero otherwise. Notice that to solve for the equation \eqref{intergral}, we should discretize  the covariance function $\sigma(t,t')$. Finally, the functional principle component scores $\hat \xi_{jk}$ can be computed by:
\begin{align}
\label{scores}
    \hat \xi_{jk} = \int_a^b (\mathcal{Y}^-_j(t)- \hat \mu(t))\hat \phi_k(t) dt
\end{align}
We can solve \eqref{scores} by a numerical integration and find the FPC-scores. These FPC-scores $\hat \xi_{jk}$ now represents each data point $j$ and they can be used for the dimensional reduction by choosing $k \ll T_0$ of them, considering that these $k$ number of FPC-socres, should explain the most of variation in the data set.   

\subsection{Step 2: Applying K-Means Clustering}
K-means algorithm is considered as a type of unsupervised learning, where we do not have any information about the quantitative or categorical response variable. After computing FPC-scores, we want to find which member of the data set can be useful for explaining the treated unit's behavior during the pre-intervention time period. To do this, I suggest using a simple clustering method proposed by \cite{kmeans} and then the data points that fall in the same cluster as the treated unit would be utilized for computing the underlying linear relation between the synthetic and the treated unit. 
In K-means clustering, the goal is to find $k$ cluster centers $\{c^1,c^2,...,c^k\}$ and assign each data point to one of clusters ($\Omega(j) \in \{1,...,k\}$) such that it minimizes the following objective function:
\begin{align}
    \min_{c,\Omega} \frac{1}{M} \sum_{j=1}^M ||\mathcal{Y'}^-_j - c^{\Omega(j)}||^2
\end{align}
where $M$ is the number of units in the data set and $\mathcal{Y'}^-$ is the reduced dimension data for the pre-intervention time period that we get from the previous step. Solving the above optimization is a NP-hard problem, so \cite{kmeans} suggested a greedy algorithm that guarantees to converge to the local minimum, but not global minimum.\\
\begin{algorithm}[t]
\SetAlgoLined
 Initialize $k$ cluster centers randomly\;
 \While{There is no changes in cluster centers}{
  Cluster assigning by: $\Omega(j) = \argmin_{l=1,...,k} ||\mathcal{Y'}^-_j - c^l||^2$\;
  Adjust the cluster centers by: $c^l = \frac{1}{|\{i:\Omega(j)=l\}|}\sum_{j:\Omega(j)=l}\mathcal{Y'}^-_j$\;
 }
 \caption{K-means Clustering}
 \label{algo1}
\end{algorithm}
As seen in Algorithm \ref{algo1}, the main idea of K-means clustering is to first initialize the cluster centers randomly and then in each iteration, assign each data point to the nearest cluster center, after which the cluster centers are recalculated. This iteration continues until there are no changes in the cluster centers. The output of this method depends on the initialization. Since its introduction there have been many studies to overcome the issue of initialization for the K-means algorithm\footnote{For a brief list of improvement on K-means algorithm, see \cite{kmeans-survey}}. 
In the case of policy analysis and synthetic control models, however, we are not dealing with big data and in the small samples that we have, the weakness of the K-means algorithm should not be an issue. Simply repeating the algorithm multiple times with different initial values and checking for the consistency of results should be sufficient to guarantee reaching the global optimum. 
In the K-means algorithm, the number of clusters is a hyperparameter that should be tuned. One simple solution for this is to use the elbow method. In the elbow method, we plot the within group sum of squares of distances from the centers, $SS(k) = \sum_{l=1}^k \sum_{j:\Omega(j)=l} ||\mathcal{Y'}^-_j - c^l||^2$, versus the cluster numbers and then choose the elbow of the curve as the number of clusters to use. Choosing the number of clusters by elbow method is subjective and in some cases there is not a clear elbow in the plot at all. A better solution to tune the number of clusters is to use Silhouettes statistics (\citet{Silhouettes}). To find the Silhouettes statistics for a specific data point, first we should compute two distances: \\
1- The distance of the data point to its own cluster center, $a(i)$ \\
2- The distance of the data point to the closest cluster center, $b(i)$  \\
Then, the Silhouettes statistics is defined as:
\begin{align*}
    s(i) = \frac{b(i) - a(i)}{\max \{a(i),b(i)\}}
\end{align*}
If the data point is in the middle of two cluster centers, we have $s(i) = 0$ and if it has a zero distance to its cluster center (meaning the data is perfectly clustered), we have $s(i) = 1$. We can calculate Silhouettes statistics for all data points for a specific number of clusters $k$. Then taking the average of them would give us the Silhouettes coefficient $SC(k)$. Based on the definition of the $s(i)$, it is obvious that the number of clusters, $k$, with the highest value of Silhouettes coefficient is the best choice for the number of clusters.

\subsection{Step 3: Robust PCA}
In the synthetic control model, we are interested in finding a linear relation between the donor pool members and the treated unit. To find such a linear relation, \cite{Amjad} use the truncated singular value decomposition to compute the low-rank structure of the donor pool, but truncated SVD can be influenced by corrupted data or outliers. To address this issue, I suggest to use the robust PCA method to retrieve the robust low-rank structure. After extracting the donor pool from the original data by the previous step, we have the matrix of data $Y_{-i} \in \mathbb{R}^{(N-1) \times T}$ (the matrix that contains the data of the outcome variable of the donor pool). We want to decompose this matrix to find the low-rank matrix $L\in \mathbb{R}^{(N-1) \times T}$ and the sparse matrix $S\in \mathbb{R}^{(N-1) \times T}$ that contains outliers and corrupted data such that:
\begin{align}
\label{decomposition}
    Y_{-i} \approx L + S
\end{align}
As discussed by \cite{R-PCA-main}, without any further assumptions on \eqref{decomposition}, decomposing a matrix to the low-rank and sparse matrices is an ill-posed problem. To be able to find a unique decomposition, \cite{R-PCA-main} considered two restrictions: first, the low-rank matrix $L$ cannot be sparse by itself and, second, the sparse matrix $S$ should not have a low-rank structure. If these two conditions are satisfied, we can solve the following optimization problem in order to find the low-rank matrix $L$ and the sparse matrix $S$: 
\begin{equation}
\begin{aligned}
\label{rpca_objective}
   & \min_{L,S} \hspace{0.5 cm} {rank(L) + \lambda ||S||_0} \\
   & \text{subject to} \hspace{0.5 cm} Y_{-i}=L+S
\end{aligned}
\end{equation}
where $||S||_0$ is the number of non-zero elements of matrix $S$ and the $\lambda$ is the trade-off between outliers or corrupted data with the low-rank matrix. The optimization problem \eqref{rpca_objective} is not tractable, but recently it has been proven that the nuclear norm is a good surrogate for the minimization of the rank of a matrix \footnote{For discussion about the necessary and sufficient conditions for using nuclear norm for the rank minimization, see \cite{nuc-norm}.} and $l1$ norm is a good surrogate to induce sparsity (\cite{l1-norm}). Therefore, to address the intractability of problem (14), \cite{RPCA-white} suggested the following convex relaxation:
\begin{equation}
\begin{aligned}
\label{rpca_objective_2}
   & \min_{L,S} \hspace{0.5 cm} {||L||_* + \lambda ||S||_1} \\
   & \text{subject to} \hspace{0.5 cm} Y_{-i}=L+S
\end{aligned}
\end{equation}
where $||L||_*$ is the nuclear norm of matrix $L$ and $||S||_1$ is the $l1$ norm of matrix $S$ which induces sparsity for the elements of $S$. $\lambda$ is the hyperparameter that needs to be tuned, but \cite{RPCA-white} showed that problem \eqref{rpca_objective_2} would converge to problem \eqref{rpca_objective}  with high probability if $\lambda =\sqrt{\frac{1}{max(N-1,T)}}$, where $N-1$ and $T$ are the dimensions of the matrix $Y$.
\par The optimization problem \eqref{rpca_objective_2} is a semidefinite programming that can be solved with the interior point method (\cite{Multiplier}), but the pre-step cost of this method which is $O(s^6)$, where $s = max(N-1,T)$, makes it impractical to use the interior point method in real-world problems.
To derive a practical and iterative method to solve problem \eqref{rpca_objective_2}, one obvious solution is to use the Lagrangian:
\begin{align}
    \mathcal{L}(S,L,\Lambda) \doteq ||L||_* + \lambda ||S||_1 + \langle \Lambda,L+S-Y_{-i} \rangle
\end{align}
where $\Lambda \in \mathbb{R}^{(N-1) \times T}$ is the matrix of Lagrange multipliers. Then the optimal solution $(S,L,\Lambda)$ is the saddle points of the Lagrangian (\cite{Wright-Ma-2021})
\begin{align}
    \sup_{\Lambda} \inf_{S,L} \mathcal{L}(S,L,\Lambda) =\sup_{\Lambda} \inf_{S,L} \hspace{0.1 cm} ||L||_* + \lambda ||S||_1 + \langle \Lambda,L+S-Y_{-i} \rangle 
\end{align}
This saddle point feature of the optimal solutions can help to generate an iterative method:
\begin{equation}
\begin{aligned} \label{lagrangian_iterative}
    & (S_{k+1},L_{k+1}) \leftarrow \argmin_{S,L} \mathcal{L}(S,L,\Lambda_k) \\ 
    & \Lambda_{k+1} \leftarrow \Lambda_{k} + \alpha_k (L_{k+1}+S_{k+1}-Y_{-i})  
\end{aligned}
\end{equation}
to solve for the optimal values of $(S,L,\Lambda)$. As \cite{augmented_lagrangian} discusses, the issue with the iterative method \eqref{lagrangian_iterative} is the chance of generating a non-feasible solution in any step, so the iterative method might fail to progress. To address the issue of generating non-feasible solutions, we can use augmented Lagrangian. The main idea behind augmented Lagrangian is to penalize the constraint more strongly. To accomplish this, we can write the problem \eqref{rpca_objective_2} as:
\begin{equation}
\begin{aligned} \label{augmented_lag}
   & \min_{L,S} \hspace{0.5 cm} {||L||_* + \lambda ||S||_1}+\frac{\mu}{2}||L+S-Y_{-i}||_F^2 \\
   & \text{subject to} \hspace{0.5 cm} Y_{-i}=L+S
\end{aligned}
\end{equation}
where $\mu>0$ is the penalty parameter and if the solution in each iteration is feasible, the extra term in problem \eqref{augmented_lag} would be zero and problem \eqref{augmented_lag} will be equivalent to problem \eqref{rpca_objective_2}. The Lagrangian for the problem \eqref{augmented_lag} is called the augmented Lagrangian and it has the form:
\begin{align}
    \mathcal{L}_{\mu}(S,L,\Lambda) \doteq ||L||_* + \lambda ||S||_1 + \langle \Lambda,L+S-Y_{-i} \rangle + \frac{\mu}{2}||L+S-Y_{-i}||_F^2
\end{align}
Similar to \eqref{lagrangian_iterative}, we can derive an iterative method for the augmented Lagrangian:
\begin{equation}
\begin{aligned} \label{augmented_lag_iter}
    & (S_{k+1},L_{k+1}) \leftarrow \argmin_{S,L} \mathcal{L}_{\mu}(S,L,\Lambda_k) \\
    & \Lambda_{k+1} \leftarrow \Lambda_{k} + \mu (L_{k+1}+S_{k+1}-Y_{-i})
\end{aligned}
\end{equation}

The iteration process of \eqref{augmented_lag_iter} is called the method of multipliers (\cite{Wright-Ma-2021}). To solve \eqref{augmented_lag_iter}, we can use the proximal gradient to find a closed form solution for each variables $S$ and $L$. 

\subsubsection{Proximal Gradient Method}
Consider the objective function that can be decomposed into two parts $g(y)+h(y)$ for $y \in R^d$, where both functions $g(y)$ and $h(y)$ are convex. Assume that the function $g(y)$ is differentiable with the Lipschitz gradient, and the function $h(y)$ is not differentiable. We know that since $g(y)$ is differentiable, the gradient descent update can be written as:
\begin{align}
    y' = y - t. \nabla g(y)
\end{align}
where $t$ is the learning rate in each step. In addition, if we assume the Hessian of $g(y)$ is equal to $\frac{I}{2}$, the quadratic approximation of $g(y)$ around the point $y$ is:
\begin{align*}
    g(z) = g(y) + \nabla g(y)^T (z-y) + \frac{1}{2t} ||z-y||_2^2
\end{align*} 
To define an iterative process over the objective function $g(y)+f(y)$, first we need to introduce the proximal operator.
\begin{definition} 
Let $h(y)$ be a convex function of $y \in R^d$, then the proximal operator over the function $h(y)$ is defined by:
\begin{align*}
    prox_{t,h} (y) = \argmin_{z \in R^d} \left(\frac{1}{2t} ||z-y||_2^2 + h(z)\right)
\end{align*} 
and for $Y_{-i} \in R^{(N-1) \times T}$, the proximal operator over the convex function $H(Y_{-i})$ is defined by:
\begin{align*}
    prox_{t,H} (Y_{-i}) = \argmin_{Z \in R^{(N-1) \times T}} \left(\frac{1}{2t} ||Z-Y_{-i}||_F^2 + H(Z)\right)
\end{align*} 
\end{definition}
Then for the case of $y \in R^d$, using the quadratic approximation for $g(y)$ and leaving the non-differentiable function $h(y)$ intact, in each iteration, we can update $y$ by:
\begin{align*}
    y' &= \argmin_{z} \left (g(y) + \nabla g(y)^T (z-y) + \frac{1}{2t} ||z-y||_2^2 + h(z) \right)\\
    &= \argmin_{z} \frac{1}{2t} ||z- (x- t \nabla g(y))||_2^2 + h(z) \\
    & = prox_{t,h} (y-t\nabla g(y))
\end{align*}
Also the generalization to the case of $Y_{-i} \in R^{(N-1) \times T}$ for the updating process above is straightforward. The idea behind the proximal gradient descent is to minimize $h(y)$ function while staying close to the point $(y- t \nabla g(y))$ at each iteration. Notice that the presence of a strong convex term $||z- (y- t \nabla g(y))||_2^2$ guarantees that $y'$ at each iteration is well-defined and unique.  Algorithm \ref{algo2} summarizes the proximal gradient descent method and as you can see, we should set $prox_t(y)$ for this algorithm to work, so the applicability of this method depends on how simple it is to compute the proximal operator for a specific problem. \cite{Boyd_Proximal} provide an extensive analysis on the convergence rate of proximal gradient descent. 
In the next section, I will discuss how the proximal operator can be used to solve the iteration process \eqref{augmented_lag_iter} for the objective function \eqref{rpca_objective_2}.

\begin{algorithm}[t]
\SetAlgoLined
 Set $prox_t(y)$\;
 Initialize $y_0$\;
 \While{not convergence}{
  Compute $y_{k+1} = prox_t(y_k - t \nabla g(y_k))$\;
 }
 \caption{Proximal Gradient Descent}
 \label{algo2}
\end{algorithm}

\subsubsection{Alternating Direction Method of Multipliers for Robust PCA}
So far I have discussed that the optimization problem \eqref{rpca_objective} can be solved with the method of multipliers which is an iteration process \eqref{augmented_lag_iter}, but minimization of $\mathcal{L}_{\mu}$ in \eqref{augmented_lag_iter} over both variables $S$ and $L$ is difficult. For the robust PCA problem \eqref{rpca_objective_2}, we can use the specific structure of the objective function to make the minimization \eqref{augmented_lag_iter} simpler. The objective function of robust PCA is separable in its variables ($L$ and $S$), so we can use a method that is known in the literature as alternating direction method of multipliers. In this method, we can use the separable structure of the objective function to update each variable in each iteration considering the other variables fixed. For the case of iterative process \eqref{augmented_lag_iter}, this means that we can update $L$ considering $S$ and $\Lambda$ fixed and repeat the same procedure for $\Lambda$ and $S$. 
Following \cite{Wright-Ma-2021}\footnote{\citet{Wright-Ma-2021} also provide an extensive proof for the convergence of ADMM under different conditions.}, we can consider sequential updates for the variable $L$:
\begin{align}
\label{L}
    L_{k+1} & = \argmin_{L} \mathcal{L}_{\mu}(S_k,L,\Lambda_k) \nonumber \\
    & = \argmin_{L} ||L||_* + \lambda ||S_k||_1 + \langle \Lambda_k,L+S_k-Y_{-i} \rangle + \frac{\mu}{2}||L+S_k-Y_{-i}||_F^2  \nonumber \\
    & = \argmin_{L} ||L||_* + \langle \Lambda_k,L+S_k-Y_{-i} \rangle + \frac{\mu}{2}||L+S_k-Y_{-i}||_F^2 \nonumber \\
    & = \argmin_{L} ||L||_* + \frac{\mu}{2} ||L+S_k-Y_{-i} + \mu^{-1}\Lambda_k||_F^2 - \frac{1}{2\mu}||\Lambda||_F^2 \nonumber \\
    & = prox_{\mu^{-1},||.||_*}\left ( Y_{-i}-S_k - \mu^{-1}\Lambda_k \right) \\
    \nonumber
\end{align}
where the proximal operator is defined over the nuclear norm, also the sequential updates for the variable $S$ can be written as:
\allowdisplaybreaks
\begin{align}
\label{S}
    S_{k+1} & = \argmin_{S} \mathcal{L}_{\mu}(S,L_{k+1},\Lambda_k)  \nonumber \\
    & = \argmin_{S} ||L||_* + \lambda ||S||_1 + \langle \Lambda_k,L_{k+1}+S-Y_{-i} \rangle + \frac{\mu}{2}||L_{k+1}+S-Y_{-i}||_F^2 \nonumber \\
    & = \argmin_{S} \lambda ||S||_1 + \langle \Lambda_k,L_{k+1}+S-Y_{-i} \rangle + \frac{\mu}{2}||L_{k+1}+S-Y_{-i}||_F^2 \nonumber \\
    & = \argmin_{S} \lambda ||S||_1 + \frac{\mu}{2} ||L_{k+1}+S-Y_{-i} + \mu^{-1}\Lambda_k||_F^2 - \frac{1}{2\mu}||\Lambda||_F^2 \nonumber \\
    & = prox_{\lambda \mu^{-1},||.||_1}\left ( Y_{-i}-L_{k+1} - \mu^{-1}\Lambda_k \right)\\
    \nonumber
\end{align}
where in this case the proximal operator is defined over the $l^1$ norm. These sequential updates over the low-rank matrix $L$ and the sparse error $S$ provide a convenient way for the update of augmented Lagrangian in \eqref{augmented_lag_iter}. We can use the following two propositions to find a closed form solution for \eqref{L} and \eqref{S}:

\begin{prop} \label{soft_thre}
Let $Y_{-i} \in R^{(N-1) \times T}$ and $y_k$ be an element of $Y_{-i}$, define $\mathcal{S}_{\tau}$ to be the elementwise soft-thresholding operator such that:
\begin{align*}
    \mathcal{S}_{\tau}(y_k)  \triangleq sign(y_k) \max(|y_k|-\tau,0)
\end{align*}
Then, we can show that $(prox_{\tau,||.||_1}(Y_{-i}))_k =  \mathcal{S}_{\tau}(y_k)$. In other words, the proximal operator over the $l^1$ norm of $Y_{-i}$, is the elementwise soft-thresholding of it. \\
Proof: \textbf{Appendix \ref{appA1}}
\end{prop}

\begin{prop} \label{SVD}
Let $Y_{-i} \in R^{(N-1) \times T}$ and $Y_{-i} = U \Sigma V^T$ be the SVD decomposition of $Y_{-i}$. Using the definition of $\mathcal{S}_{\tau}$, we can define the singular value thresholding operator by:
\begin{align*}
    \mathcal{D}_{\tau}(Y_{-i}) = U \mathcal{S}_{\tau}(\Sigma) V^T
\end{align*}
where $\mathcal{S}_{\tau}$ is the elementwise soft-thresholding operator. Then, we can show that $prox_{\tau,||.||_*}(Y_{-i}) = U \mathcal{S}_{\tau}(\Sigma) V^T$. In other words, the proximal operator over the nuclear norm of $Y_{-i}$, is the elementwise soft-thresholding over its singular values. \\
Proof: \textbf{Appendix \ref{appA2}}
\end{prop}
Based on the two propositions above, we can find a closed form solution for proximal operator of \eqref{L} and \eqref{S}:
\begin{align} \label{L_last}
    L_{k+1} & = \argmin_{L} \mathcal{L}_{\mu}(S_k,L,\Lambda_k) \nonumber  \\
    & = prox_{\mu^{-1},||.||_*}\left ( Y_{-i}-S_k - \mu^{-1}\Lambda_k \right) \nonumber\\
    & = \mathcal{D}_{\mu^{-1}} \left( Y_{-i}-S_k - \mu^{-1}\Lambda_k \right)
\end{align}
\begin{align} \label{S_last}
    S_{k+1} & = \argmin_{S} \mathcal{L}_{\mu}(S,L_{k+1},\Lambda_k)  \nonumber \\
    & = prox_{\lambda \mu^{-1},||.||_1}\left ( Y_{-i}-L_{k+1} - \mu^{-1}\Lambda_k \right) \nonumber\\
    & = \mathcal{S}_{\lambda \mu^{-1}} \left(  Y_{-i}-L_{k+1} - \mu^{-1}\Lambda_k \right)
\end{align}
Putting all these results together, we can solve robust PCA problem by ADMM. In this method (Algorithm \ref{algo3}), we first minimize $\mathcal{L}_{\mu}$ with respect to $L$ by \eqref{L_last}, considering all other variables fixed. Then, we minimize $\mathcal{L}_{\mu}$ with respect to $S$ by \eqref{S}, again all other variables fixed. At the end, ADMM updates the dual variable $\Lambda$ according to \eqref{augmented_lag_iter}. Now that we have discussed an algorithm to compute the robust PCA, in the next section, we can use all the previous steps (step 1, 2 and 3) to introduce robust PCA synthetic control.

\begin{algorithm}[t]
\SetAlgoLined
 Initialize $S_0$, $\Lambda_0$, $\mu >0$\;
 \While{not convergence}{
  Update $L_{k+1}$ by $L_{k+1}= \mathcal{D}_{\mu^{-1}} \left( Y_{-i}-S_k - \mu^{-1}\Lambda_k \right)$\;
  Update $S_{k+1}$ by $S_{k+1}= \mathcal{S}_{\lambda \mu^{-1}} \left(Y_{-i}-L_{k+1} - \mu^{-1}\Lambda_k \right)$\;
  Update $\Lambda_{k+1}$ by $\Lambda_{k+1} = \Lambda_{k} + \mu (L_{k+1}+S_{k+1}-Y_{-i})$\;
 }
 \caption{Robust PCA by ADMM}
 \label{algo3}
\end{algorithm}

\subsection{Robust PCA Synthetic Control}
\label{sec_2.4}
Synthetic control model is a method for counterfactual estimation. Consider a treated unit $Y_{i}$. This unit would receive a treatment at time $T_{0}$ and we can observe the output of unit $Y_{i}$ both before receiving the treat $1,...,T_{0}$ and after the treat $T_{0}+1,...,T$ where $T$ is the last observed time period. The idea behind the synthetic control model is to find a donor pool, denoted by $Y_{-i}$, that can explain the behavior of the treatment unit before receiving the treat, the best. Using this donor pool, we can find a relation between the treated unit and the donor pool for the time period before the treatment. This relation can then be used to estimate the counterfactual of treated unit after receiving the treat.\\
In the original synthetic control model proposed by \cite{abadie2003}, we need what \cite{Amjad} call ``expert in the field''. This means that to implement the synthetic control model, we need experts to find the appropriate donor pool based on the similarity between the treated unit and the donor pool members. Besides that, we need to consider suitable features (explanatory variables of the outcome variable) to implement the original synthetic control model. Choosing donor pool members and covariates can change the output of counterfactual estimation significantly. Here, I use machine learning techniques to overcome these issues.\\
The method I suggest here, called robust PCA synthetic control, is an intuitive data-driven solution to address the issues of classic synthetic control model. First, we need to choose the donor pool for the treated unit. For this, first we can use the functional principle components to reduce the dimensionality of a data set and overcome the potential curse of dimensionality in K-means algorithm. After computing principle component scores for the pre-treatment period of a whole data set (including treated and non-treated units), we have the reduced dimension presentation of each data point. Then we can use the K-means algorithm to cluster these data points. The data points that fall in the same cluster as the treated unit have the closest distance to it and we can use them as the donor pool.\\
Now that we have the appropriate donor pool, we can extract the low-rank structure of non-treated units $Y_{-i}$ for the pre-treatment area. This approach is inspired by \cite{Amjad}, but in their work, they use thresholding on singular value decomposition to find this low-rank structure and as \citet{PCA_issue} mentioned, SVD is sensitive to outliers or missing data, so I use the robust PCA algorithm to solve this issue.\\
The output of the robust PCA would give us the low-rank structure of the donor pool ($L$). I denote the pre-intervention part of $L$ by $L^{-}$, which we can then use to estimate the linear relation between the treated unit and non-treated units for the pre-intervention period by a simple least square method:
\begin{align}
    &\hat{\beta} = \argmin_{z \in R^{N-1}} ||Y_{i}^{-} - (L^{-})^Tz||_2^2 \\
     &\text{subject to} \hspace{0.4 cm} z \geq 0 \nonumber
\end{align}
where I assume there are $N$ units (including treated and donor pool members), so $Y_{-i} \in R^{N-1 \times T}$ represents $N-1$ members of non-treated units for the $1,...,T$ time span. I also impose the constraint of positive relation between the donor pool and the treated unit. This is intuitive, as I expect that clustering on FPC-scores would put the unit members with a positive relation with the treated unit in the same cluster. After computing $\hat \beta$ values, we can estimate the counterfactual of a treatment unit as:
\begin{align}
    \hat Y_{i}^{+} = (L^{+})^T \hat \beta
\end{align}
where $L^{+}$ denotes the post intervention part of the low-rank matrix $L$. \\
There are five hyperparameters in the robust PCA synthetic control algorithm: one is the number of FPC-scores, second is the number of clusters in the K-means algorithm, and the others are related to the computation of robust PCA. To choose the number of FPC-scores, we can compute the proportion of the explained variation in the data for the specific number of FPC-scores and then pick the number of FPC-scores that can explain the most variation in the data\footnote{As a rule of thumb, we should choose the FPC-score that can explain at least $95$ percent of the variation in the data.}. For the hyperparameter of K-means algorithm, as I discussed before, we can use the Silhouette statistics. For the hyperparameter $\lambda$ in robust PCA, I followed the recommendation of \cite{RPCA-white} for the convergence of the algorithm and based on the dimensions of $Y_{-i}$, I considered $\lambda =\sqrt{\frac{1}{max(N-1,T)}}$. For the other two hyperparameters in robust PCA, $\tau$ and $\mu$, I considered the recommended values by \cite{data_Driven} which are $\tau = 10^{-7} \times||Y_{-i}||_F$ and $\mu = \frac{(N-1)\times T}{4\times \sum|vec(Y_{-i})|}$ where $vec(.)$ is the vectorization operator. Notice that these values for the hyperparameters are merely rule of thumb and one could use methods like cross validation over the pre-intervention period to tune these hyperparameters. Finally, I summarize the robust PCA synthetic control algorithm in algorithm \ref{algo4}. In the next section, I will implement the robust PCA synthetic control on the case of West Germany reunification with the mentioned hyperparameters.

\begin{algorithm}[]
\SetAlgoLined
  \textbf{Step 1.}
  \begin{itemize}
      \item Define $\mathcal{Y}^{-}$ as the pre-intervention period for both treated unit and \\ non-treated units in the data set
      \item Compute  FPC-scores $\xi$ for all units that explain most of variation in the data
  \end{itemize}
  \textbf{Step 2.}
  \begin{itemize}
      \item Apply K-means algorithm on FPC-scores and extract the donor pool $Y_{-i}$
  \end{itemize}
  \textbf{Step 3.}
  \begin{itemize}
      \item Run Robust PCA by ADMM over the donor pool $Y_{-i}$ and compute \\ $L$ and $S$
  \end{itemize}
  \textbf{Step 4.}
  \begin{itemize}
      \item Compute the relation between the treated unit and the donor pool by:
      \begin{align}
      &\hat{\beta} = \argmin_{z \in R^{N-1}} ||Y_{i}^{-} - (L^{-})^Tz||_2^2 \nonumber \\
     &\text{subject to} \hspace{0.4 cm} z \geq 0 \nonumber
  \end{align}
  \end{itemize}
  \textbf{Step 5.}
  \begin{itemize}
      \item Estimate the counterfactual of the treated unit by:
      \begin{align*}
          \hat Y_{i}^{+} = (L^{+})^T \hat \beta
      \end{align*}
  \end{itemize}
 \caption{Robust PCA Synthetic Control}
 \label{algo4}
\end{algorithm}

\section{Empirical Study}
\label{sec3}
\begin{figure}[]
\caption{\textbf{FPC-scores}}
\centering
\includegraphics[width=7cm,height =5 cm]{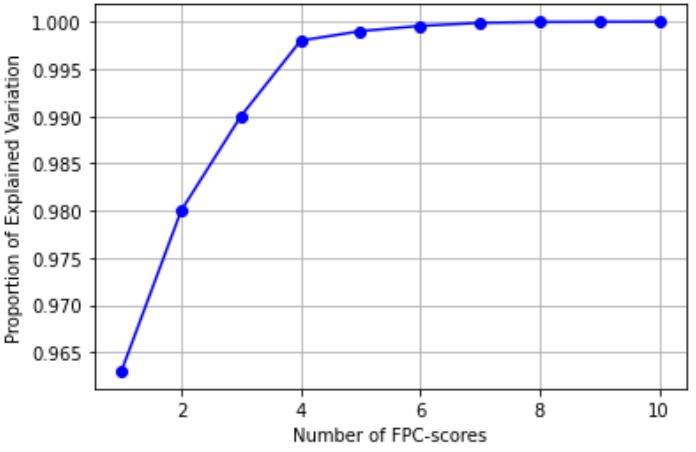}
\label{fig1}
\end{figure}

\begin{figure}
\caption{Elbow method (a) and Silhouettes coefficient (b) for tuning the number of clusters in K-means algorithm.}
\centering
\begin{subfigure}{.5\textwidth}
   \caption{\textbf{Elbow Method}}
  \centering
  \includegraphics[width=7cm,height =5 cm]{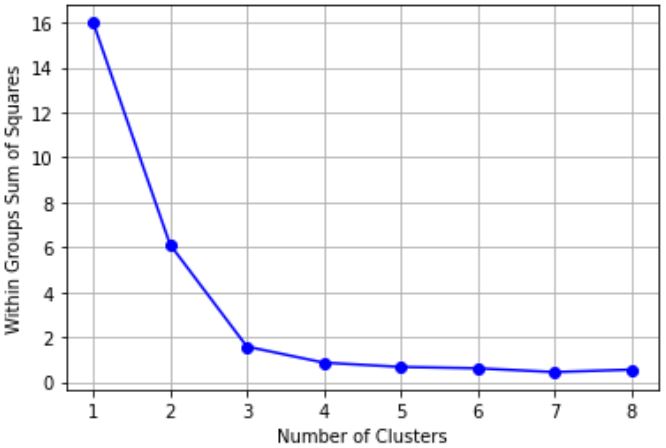}
  \label{fig2}
\end{subfigure}%
\begin{subfigure}{.5\textwidth}
  \caption{\textbf{Silhouettes Coefficient}}
  \centering
  \includegraphics[width=7cm,height =5 cm]{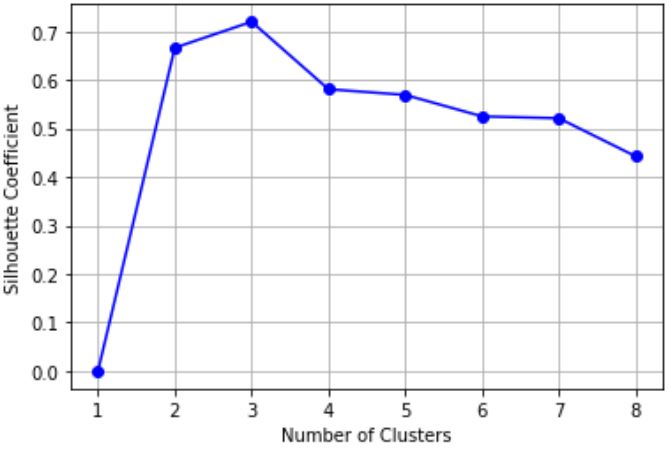}
  \label{fig3}
\end{subfigure}
\label{wholefig2}
\end{figure}

In this section, I apply the robust PCA synthetic control algorithm to the case of West Germany reunification. To make all conditions similar to \cite{abadie2015}, I used the same country-level panel data of $16$ OECD members from $1960-2003$. In general, to implement the robust PCA synthetic control algorithm, we do not need to find an appropriate donor pool as the algorithm would take care of that, but here I used the same data set of \cite{abadie2015} just for an equitable comparison.\\
In $1990$, West Germany reunified with East Germany. \cite{abadie2015} use the synthetic control method to answer this question: what would have been the per capita GDP of West Germany if this reunification never happened?\\
To implement the robust PCA synthetic control, first I computed the FPC-scores of the $16$ OECD members in addition to West Germany for the pre-intervention period of $1960-1990$. Figure \ref{fig1} shows the percentage of explained variation in the data based on the number of FPC-scores. As you can see, even considering the first FPC-score can explain more than $95 \%$ of the variation in the data. So, to reduce the dimensionality of data set, I considered the first FPC-score for the next step of my method, which is applying K-means algorithm \footnote{The outcome of the k-means algorithm does not change even when selecting up to $5$ FPC-scores.}.  \\
For K-means algorithm, we first need to find the optimized number of clusters. To do this, I use both elbow method and Silhouettes coefficient (figure \ref{fig2} and figure \ref{fig3}). In figure \ref{fig3}, the Silhouettes coefficient would reach its maximum when the data has been clustered by $3$ groups. We can have the same conclusion by looking at the figure \ref{fig2} where the elbow of the plot corresponds to the $3$ clusters.

\begin{table}[]
\centering
\caption{\textbf{K-means Algorithm with $k=3$}}
{%
\begin{tabular}{lll}
\hline
Cluster 1    & Cluster 2 & Cluster 3   \\
\hline \hline 
United Kingdom           & Greece    & United States         \\
Belgium      & Portugal  & Switzerland \\
Denmark      & Spain     &             \\
France       &           &             \\
\textbf{West Germany} &           &             \\
Italy        &           &             \\
Netherlands  &           &             \\
Norway       &           &             \\
Japan        &           &             \\
Australia    &           &             \\
New Zealand  &           &             \\
Austria      &           &           \\
\hline
\end{tabular}%
}
\label{tab1}
\end{table}

Table \ref{tab1} shows the result of K-means algorithm on the first FPC-score of the data set. Out of $16$ countries, $11$ counties fall in the same cluster of West Germany. I use these countries as the donor pool of West Germany for the post-intervention estimation. \\
\interfootnotelinepenalty=10000
Now that we have the donor pool, we can run step 3 and 4 of algorithm \ref{algo4} to find the synthetic weights. Table \ref{tab2} shows the synthetic weights of robust PCA, along with the weights of synthetic control model (reported in \cite{abadie2015}) and weights of robust synthetic control model (suggested in \cite{Amjad}) for the West Germany \footnote{To implement the robust synthetic control model and extract the low-rank structure of the donor pool, I considered the first $2$ singular values of the donor pool. In appendix \ref{appB}, I explained the justification behind it.}. The robust PCA synthetic control considers Austria, France, New Zealand and Norway as the synthetic of West Germany, while the synthetic control considers Austria, Japan, Netherlands, Switzerland and United Sates. In contrast with these two methods, the robust synthetic control puts positive weights on all countries in the donor pool and this can create issues specially in the presence of outliers. 

\par All three models consider a linear relation between the treated unit and its synthetics, but there are some differences in their assumptions. In robust PCA, the weights are conditioned to be greater or equal to zero. As I mentioned before, the idea behind this assumption is that the step 1 and 2 of robust PCA algorithm would pick up members of the donor pool with similar behavior as the treated unit during the pre-intervention period. In \cite{abadie2015}, the weights are conditioned to be between zero and one and their sum should be one. In other words, in the synthetic control, the weights are the convex hull of the treated unit. Robust synthetic control, on the other hand, does not impose any constraint on the synthetic weights.

\begin{table}[]
\centering
\caption{\textbf{Synthetic Weights for West Germany}}

\resizebox{\textwidth}{!}{%
\begin{tabular}{llll}
\hline
Country        & \begin{tabular}[c]{@{}l@{}}Robust PCA\\ Synthetic Control\\ Weight\end{tabular} & \begin{tabular}[c]{@{}l@{}}Synthetic Control\\ Weight\end{tabular} & \begin{tabular}[c]{@{}l@{}}Robust \\ Synthetic Control\\ Weight\end{tabular} \\ \hline \hline
Australia      & 0                                                                               & 0                                                                                       & 0.07                                                                         \\
Austria        & 0.02                                                                            & 0.42                                                                                    & 0.08                                                                         \\
Belgium        & 0                                                                               & 0                                                                                       & 0.08                                                                         \\
Denmark        & 0                                                                               & 0                                                                                       & 0.06                                                                         \\
France         & 0.35                                                                            & 0                                                                                       & 0.07                                                                         \\
Greece         & -                                                                               & 0                                                                                       & 0.05                                                                         \\
Italy          & 0                                                                               & 0                                                                                       & 0.08                                                                         \\
Japan          & 0                                                                               & 0.16                                                                                    & 0.08                                                                         \\
Netherlands    & 0                                                                               & 0.09                                                                                    & 0.07                                                                         \\
New Zealand    & 0.29                                                                            & 0                                                                                       & 0.05                                                                         \\
Norway         & 0.48                                                                            & 0                                                                                       & 0.08                                                                         \\
Portugal       & -                                                                               & 0                                                                                       & 0.04                                                                         \\
Spain          & -                                                                               & 0                                                                                       & 0.05                                                                         \\
Switzerland    & -                                                                               & 0.11                                                                                    & 0.08                                                                         \\
United Kingdom & 0                                                                               & 0                                                                                       & 0.06                                                                         \\
United States  & -                                                                               & 0.22                                                                                    & 0.09           \\ \hline                                                             
\end{tabular}%
}

\begin{tablenotes}
      \small
      \item Notes: The ``-'' indicates that the robust PCA method does not use this country as the donor pool of West Germany
    \end{tablenotes}
\label{tab2}
\end{table}

\begin{figure}[]
\caption{\textbf{Trends in Per Capita GDP: West Germany versus Synthetic West Germany}}
\centering
\includegraphics[width=\textwidth,height =8 cm]{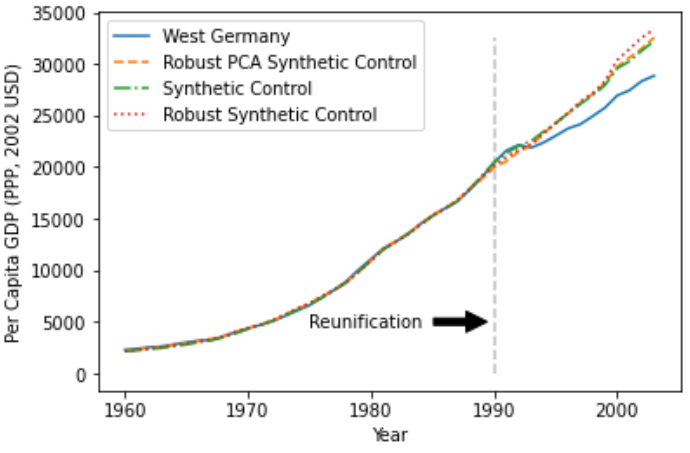}
\label{fig4}
\end{figure}

\begin{figure}[]
\caption{\textbf{Per Capita GDP Gap Between West Germany and Synthetic West Germany}}
\centering
\includegraphics[width=\textwidth,height =8 cm]{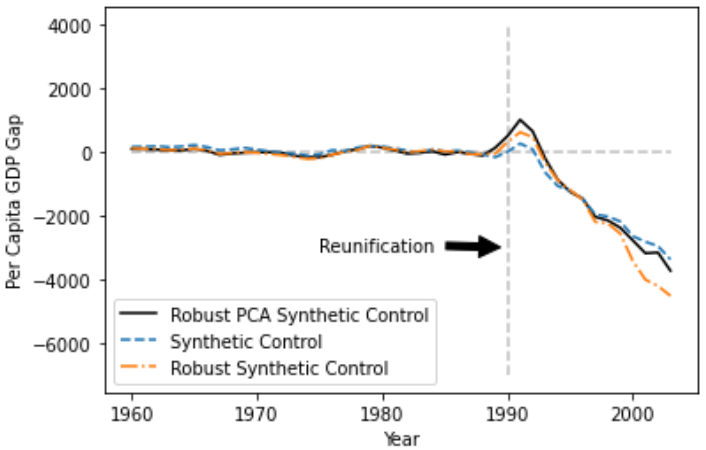}
\label{fig5}
\end{figure}

\noindent Given the synthetic weights for West Germany for all three models, we can estimate the counterfactual of the treated unit for the post intervention period. Figure \ref{fig4} shows the estimation of the counterfactual of West Germany after year 1990 for all models, along with its actual per capita GDP. The difference between the estimation of these models is better displayed in figure \ref{fig5}, where the gap between actual per capita GDP of West Germany and the estimation of each model is plotted. For the first two years of reunification (1991 and 1992), all three models (more pronounced in the robust PCA) show an increase in the per capita GDP of West Germany due to the initial increase in the demand. After that, all three models suggest a decrease in the per capita GDP of West Germany compared to its counterfactual estimation. The robust PCA synthetic control stays closer to the synthetic control, while the robust synthetic control shows a higher difference in West Germany and its counterfactual.

\begin{figure}[]
\caption{\textbf{Placebo Reunification 1975–Trends in Per Capita GDP: West Germany versus Synthetic West Germany}}
\centering
\includegraphics[width=\textwidth,height =7.5 cm]{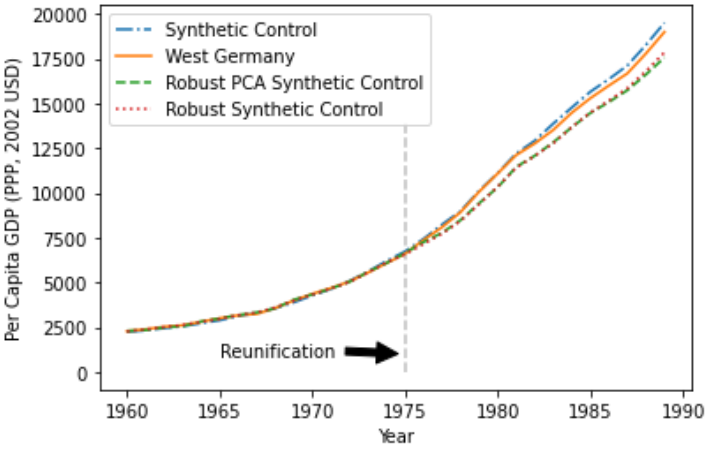}
\label{fig6}
\end{figure}

\noindent To assess how reliable the counterfactual estimations of these models are, I implement placebo studies in two ways. First placebo study in time: In this method, we consider an earlier time before the actual intervention, for example the year $1975$. Then we estimate the counterfactual of the treated unit from $1975$ up to the actual intervention time, $1990$ for the case of West Germany. Figure \ref{fig6} shows the result of placebo study for the the models, along with the actual per capita GDP of West Germany between $1976$ to $1990$. The synthetic control model slightly overestimates the per capita GDP, while the robust PCA synthetic control and robust synthetic control models both slightly underestimate it. Although the synthetic control model more closely parallels the actual per capita GDP, we should also consider the fact that synthetic control in \citet{abadie2015} utilizes $5$ variables for the counter factual estimation, whereas the robust PCA synthetic control and robust synthetic control use only the outcome variable (i.e. per capita GDP) for this estimation. Crucially, the counterfactual estimation of all these models do not diverge from the actual per capita GDP of West Germany and successfully capture its trend from $1975$ onward. 

\begin{figure}[]
\caption{\textbf{Ratio of Post-Reunification RMSPE to Pre-Reunification RMSPE: West Germany and Control Countries}}
\centering
\includegraphics[width=\textwidth,height = 19.5 cm]{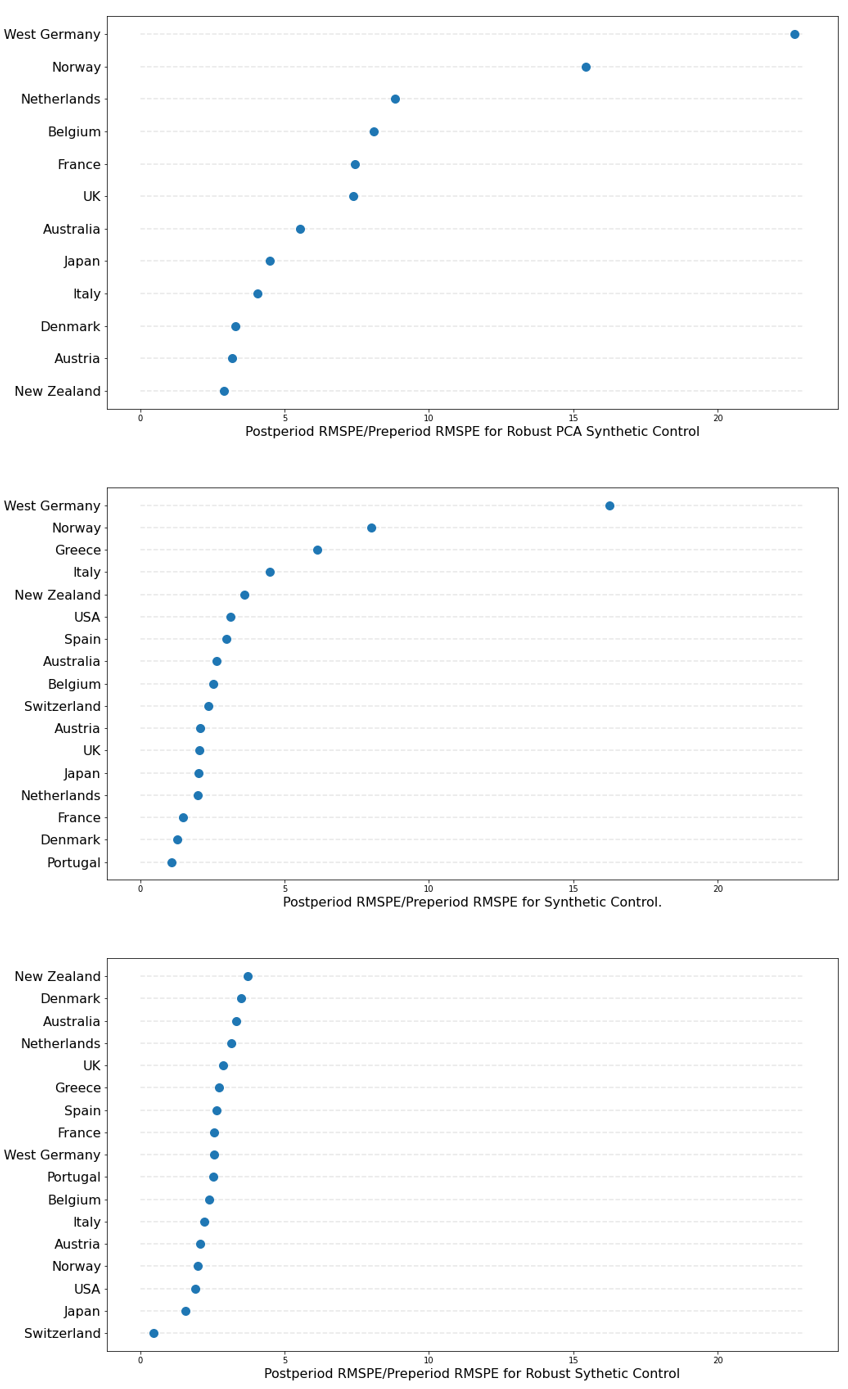}
\label{fig7}
\end{figure}

\noindent The second method of placebo study is to reassign the treatment to other units. In this form of placebo study, we iteratively assume that the intervention happened to one of the units in the donor pool and we use our model to estimate the counterfactual of the treated unit. Therefore, for each unit in the donor pool, we would have model fittings for the pre-intervention period and counterfactual estimations for the post-intervention time. Then, we can use $RMSPE_i$, which is define for each unit $i$ as:
\begin{align*}
    RMSPE_i = \left( \frac{1}{T} \sum_{t}^T \left ( Y_{it}-\sum_{j \neq i} \hat{\beta_j} Y_{jt}\right )^2 \right )^{1/2}
\end{align*}
To find the lack of goodness of fit for each unit $i$ and for both pre-intervention period $1,...,T_0$ and post-intervention period $T_0,...,T$. The idea is that if the actual treated unit, West Germany, experienced a significant difference in its per capita GDP as the result of intervention, the ratio of its post-intervention $RMSPE$ to its pre-intervention $RMSPE$ should be significantly higher than other units in the donor pool which did not receive the actual intervention. Figure \ref{fig7} shows the ratio of post-reunification $RMSPE$ to the pre-unification $RMSPE$ for all units of the donor pool for all three models. This figure demonstrates that this ratio is noticeably higher for West Germany compared to other units in the donor pool when we apply the robust PCA synthetic control and the synthetic control model suggested in \cite{abadie2015}. Interestingly, the robust synthetic control fails to satisfy this placebo study as the ratio of two $RMSPE$, for the case of West Germany is not higher compared to the other countries. As previously mentioned, one possible explanation for the failure of robust synthetic control could be that the underlying process of this method is vulnerable to the effects of outliers, and the data set may contain outliers.

\begin{figure}[]
\caption{\textbf{Leave-One-Out Distribution of the Synthetic Control for West Germany. Top graph: Robust PCA Synthetic Control; middle graph: Abadie et al. 2015; bottom graph: Robust Synthetic Control}}
\centering
\includegraphics[width=\textwidth,height = 19.5 cm]{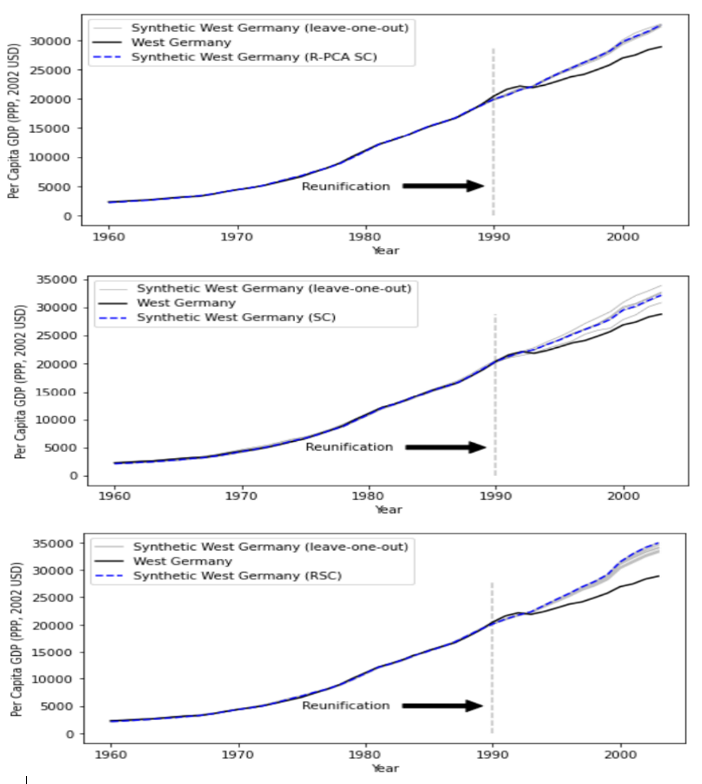}
\label{fig8}
\end{figure}

\noindent Finally, to show that the findings of robust PCA synthetic control are not sensitive to the weight of a specific country in the synthetic, I follow the \cite{abadie2015} to run a similar robustness test. As previously discussed, to estimate the counterfactual of West Germany by robust PCA synthetic control, five countries (Austria, France, New Zealand and Norway) have a positive weight, and for the synthetic control model these five countries are Austria, Japan, Netherlands, Switzerland and United Sates, while the robust synthetic control model puts positive weights on all countries. Our counterfactual estimation is robust if removing one of the synthetic members does not change the conclusion about the impact of reunification on West Germany, also ideally should not change the counterfactual estimation drastically. Figure \ref{fig8} shows the result of the robustness test on all three models. For the robust PCA synthetic control and synthetic control models, in each iteration, I drop one of the five countries with the positive synthetic weight and I re-estimate the counterfactual (gray lines). Robust PCA synthetic control shows a robust behavior for the estimation of counterfactual in the absent of one synthetic member. In the case of synthetic control, when the US is eliminated from the donor pool, the counterfactual estimation gets very close to the actual per capita GDP of West Germany or even cut it in some years. For the robust synthetic control, I drop all members of donor pool one by one in each iteration, because all countries in the donor pool have a positive weight for the counterfactual estimation and the result shows that this method can pass the robustness test. In general, the robust PCA synthetic control shows the smallest fluctuation around its counterfactual estimation in the absent of one of its synthetic units and it has the best performance for the robustness test.

\section{Simulation Study}
\label{sec4}
In this section, I conduct a simulation study to analyze the performance of robust PCA synthetic control. To do so, I first generate two different processes with additive noise:
\begin{align}
\label{functions}
     & f_1(t) = 0.3.(t \; mod(T+1)) - (t  \; mod(10)). sin(\frac{t}{\pi}) + (t \; mod(10)). cos(\frac{t}{\pi}) + \epsilon_t \nonumber \\
        & f_2(t) = log(t) + 4.sin(\frac{t}{\pi}) + 4.cos(\frac{t}{\pi}) + \epsilon_t
\end{align}
where $\epsilon_t$ is an i.i.d Gaussian noise with a mean of zero and variances of 1, 4, 9, 16, and 25. For each specific value of the variance, I generate $N_1 =100$ of $f_1(t)$ and  $N_2=100$ of $f_2(t)$ where $t \in [0,...,T=250]$. I assume that the intervention happens at $t = 150$.
\begin{figure}[h]
\caption{Generated data for $\sigma^2 = 25$, along with the mean of the underlying process}
\includegraphics[width=\textwidth]{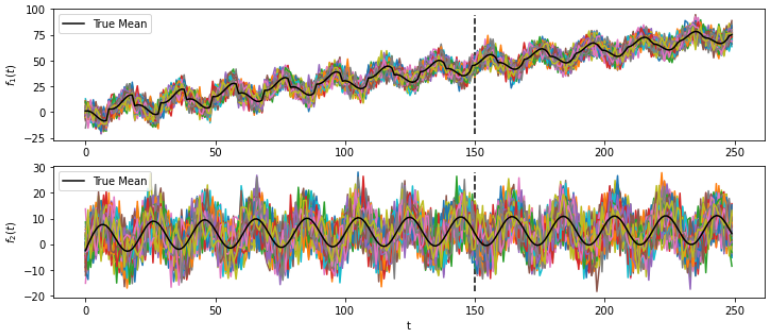}
\label{data_generated}
\end{figure}

\begin{figure}[]
\caption{Proportion of explained variance by FPC-scores (left) and  Silhouettes statistics for the number of clusters (right), for the case of $\sigma^2 = 25$}
\includegraphics[width=\textwidth,height = 5 cm]{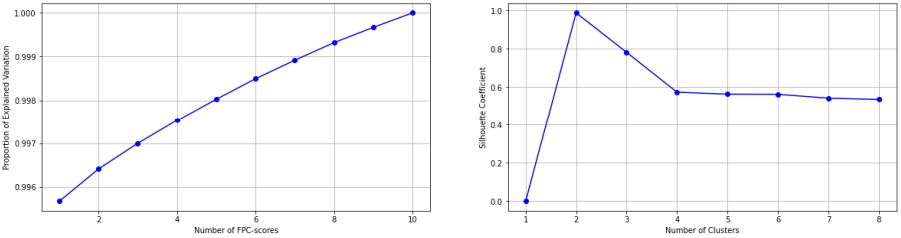}
\label{fpc_sil_sim}
\end{figure}

Figure \ref{data_generated} shows the plot of the two processes (\ref{functions}) along with their true underlying mean for the case of the highest noise variance $\sigma^2 = 25$. I implement the functional principal component analysis over the pre-invervention time period $t=1,...,150$ and then cluster these reduced dimension data points using K-means algorithm. For all cases of variance of 1, 4, 9, 16, and 25, the first FPC-scores can explain more than $95 \%$ of the variation in the data and Silhouettes statistics (Figure \ref{fpc_sil_sim})\footnote{Figure \ref{fpc_sil_sim} shows the explained variation by FPC-scores and Silhouettes statistics for the number of clusters, just for the case of $\sigma^2 = 25$, but the conclusion would not change for the other values of variances.} can determine the two clusters of the data points which are exactly the two underlying processes in (\ref{functions}). 

The accuracy of the clustering for the underlying processes (\ref{functions}), even with one FPC-score, is $100 \%$ and the K-means can differentiate between these two processes for all amount of noises. Then, I discard $f_2(t)$ and I continue the analysis over $f_1(t)$ because this process has a more complicated structure.\\
For $f_1(t)$, assume that the true means is the treatment unit that we want to estimate and the noisy data are the donor pool of it. I use the robust PCA to extract the low-rank structure of the donor pool. Then I estimate the relation between the true mean and the low-rank structure of the donor pool for the pre-intervention period.  
\begin{figure}[]
\caption{Noisy observations (gray), true means (blue), and the estimates
from the algorithm for $\sigma^2 = 1,4,9,16,25$ (red)}
\includegraphics[width=\textwidth,height= 8 cm]{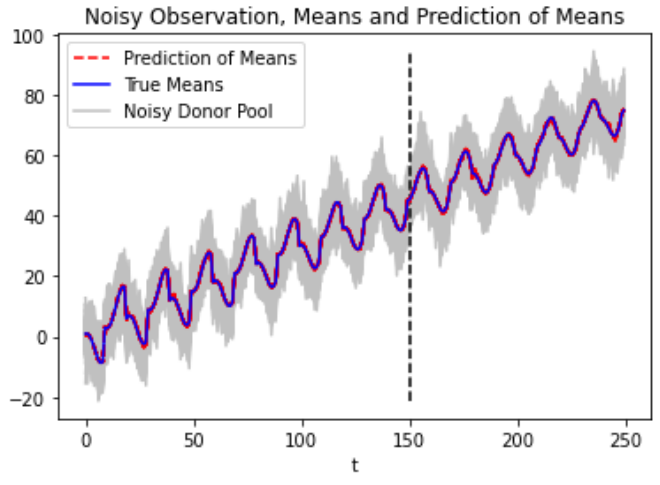}
\label{sim_estimation}
\end{figure}

\begin{table}[]
\centering
\caption{\textbf{Pre-invervention vs. Post-invervention Error}}
\resizebox{!}{!}
{%
\begin{tabular}{lll}
\hline
Noise    & Pre-invervention error & Post-invervention error  \\
\hline \hline 
1          & 0.09    & 0.13         \\
4          & 0.19  & 0.25 \\
9          & 0.29     &  0.38           \\
16           &  0.39         & 0.51            \\
25          & 0.49           &  0.64           \\

\hline
\end{tabular}%
}
\label{tab3}
\end{table}

\begin{figure}[]
\caption{True means (blue) and the estimates
from the algorithm for $\sigma^2 = 1,4,9,16,25$ (red) with $30\%$ of randomly missing data}
\includegraphics[width=\textwidth,height= 8 cm]{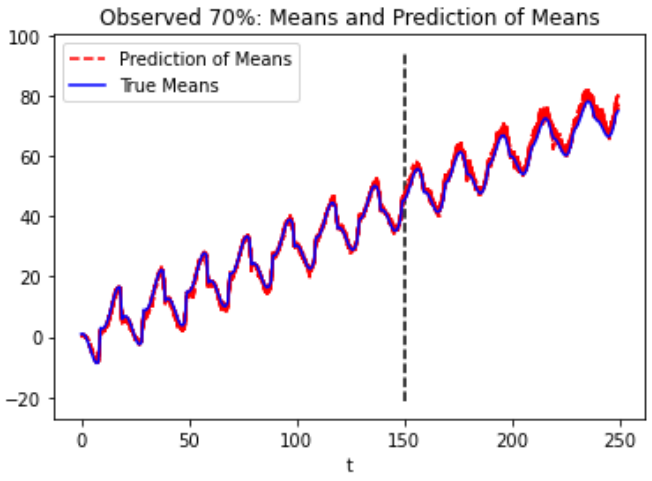}
\label{sim_missing}
\end{figure}

\begin{table}[]
\centering
\caption{\textbf{Pre-invervention vs. Post-invervention Error with Missing Data }}
\resizebox{!}{!}
{%
\begin{tabular}{lll}
\hline
Noise    & Pre-invervention error & Post-invervention error \\
\hline \hline 
1          & 0.27    & 0.65        \\
4          & 0.52  & 1.14 \\
9          & 0.86     &  1.69          \\
16           &  1.07         & 2.65            \\
25          & 1.36          &  2.59           \\

\hline
\end{tabular}%
}
\label{tab4}
\end{table}

Figure (\ref{sim_estimation}) shows the estimation of the true means for both pre-invervention and post-intervention period for all values of variances of 1, 4, 9, 16, and 25. Although the shape of the true mean has changed slightly after the intervention, robust PCA synthetic control remains close to the true mean even with a high amount of noise in data. Table \ref{tab3} reports the RMSPE of estimation for both pre-invervention and post-invervention period, showing that these errors stay close to each other for different amounts of noise.\\
In the next step, I repeat this procedure but this time dropping $30 \%$ of the data randomly. The estimation of the true means become less accurate with $30 \%$ missing data (Figure (\ref{sim_estimation}) and Table \ref{tab4}), but robust PCA synthetic control remains fairly close to the true means after the intervention. Note that the synthetic control model does not work in the presence of the missing data, but robust PCA synthetic control can estimate the counterfactual even with relatively high volume of missing data and noise.

\section{Conclusion}
\label{sec5}
Synthetic control model has recently became a popular method for counterfactual estimation, although some obstacles remain. In this study, I suggest a five-step algorithm, robust PCA synthetic control, to overcome some of these issues. Robust PCA synthetic control uses the combination of functional principal component analysis and K-means algorithm to select the donor pool, so the selection of donor pool would no longer be subjective. Then, this method uses the robust principle components analysis to find the low-rank structure of the donor pool. After finding the low-rank structure, robust PCA synthetic control estimates a linear relation between the treated unit and the donor pool over the pre-intervention period. By this relation we can find the counterfactual estimation of the treated unit for post-intervention period. \\
In addition to finding the donor pool, robust PCA synthetic control model can estimate the counterfactual just based on the outcome variable and it does not need an expert to figure out the relevant features (covariates) of the treated unit.\\
Finally, I implement robust PCA synthetic control method on the case of West Germany reunification with the East Germany on $1990$ to estimate the counterfactual of West Germany's per capita GDP. I document that my method can outperform the robust synthetic control suggested by \cite{Amjad} in placebo studies and it is less sensitive to the weights of synthetic members compare to synthetic control model implemented in \cite{abadie2015}.

\newpage
\begin{appendices}
\section{Propositions Proof}
\subsection{Proof of Proposition \ref{soft_thre}}
\label{appA1}
\begin{proof}
To proof the proposition, consider the definition of proximal operator:
\begin{align*}
    prox_{\tau,||.||_1} (Y_{-i}) &= \argmin_{Z \in R^{(N-1) \times T}} \left(\frac{1}{2t} ||Z-Y_{-i}||_F^2 + ||Z||_1\right) \\
    &= \argmin_{Z \in R^{(N-1) \times T}} \left(\frac{1}{2} ||Z-Y_{-i}||_F^2 + \tau||Z||_1\right)
\end{align*}
The above optimization would reach its minimum when the subgradients of $\frac{1}{2} ||Z-Y_{-i}||_F^2 + \tau||Z||_1$ contains zero:
\begin{singlespace}
\begin{equation*}
    0 \in (Z-Y_{-i}) + \tau \partial ||Z||_1 = \begin{cases}
      z_k - y_k + \tau, & z_k>0 \\
      -y_k + \tau [-1,1], &  z_k=0\\
      z_k - y_k - \tau, & z_k<0 
    \end{cases}
\end{equation*}
\end{singlespace}
which means that:
\begin{align*}
    (prox_{\tau,||.||_1}(Y_{-i}))_k = sign(y_k) \max(|y_k|-\tau,0) =\mathcal{S}_{\tau}(y_k)
\end{align*}
\end{proof}

\subsection{Proof of Proposition \ref{SVD}}
\label{appA2}
\begin{proof}
Similar to the previous proof, we have the proximal operator over the nuclear norm as:
\begin{align*}
    prox_{\tau,||.||_*} (Y_{-i}) &= \argmin_{Z \in R^{(N-1) \times T}} \left(\frac{1}{2t} ||Z-Y_{-i}||_F^2 + ||Z||_*\right) \\
    &= \argmin_{Z \in R^{(N-1) \times T}} \left(\frac{1}{2} ||Z-Y_{-i}||_F^2 + \tau||Z||_*\right)
\end{align*}
Following \citet{proof}, $\hat{Z} = \mathcal{D}_{\tau}(Y_{-i})$ is the solution of the above optimization if:
\begin{align}
    0 \in (\hat{Z}-Y_{-i}) + \tau \partial ||Z||_*
\end{align}
We know that the subgradients of nuclear norm is (see \citet{nuclear_norm}):
\begin{align}
\label{subgrad}
    \partial ||Z||_* = \{UV^* + W: W \in R^{(N-1) \times T}, U^*W = 0, WV = 0, ||W||_2 \leq 1\}
\end{align}
Using the SVD decomposition, we can write $\hat{Z}$ and $Y_{-i}$ as:
\begin{align*}
    & Y_{-i} = U_1 \Sigma_1 V_1^* + U_0 \Sigma_0 V_0^* \\
    & \hat{Z} = U_1(\Sigma_1 - \tau I)V_1^*
\end{align*}
where the $U_1,V_1$ are the left and right (respectively) singular vectors of $Y_{-i}$ related to the singular values greater than $\tau$, while the $U_0,V_0$ are related to the singular values smaller or equal to $\tau$. Based on the SVD decomposition, we have:
\begin{align*}
    \hat{Z} - Y_{-i} = - \tau (U_1V_1^* + W), \hspace{0.5 cm} W = \tau^{-1}U_0\Sigma_0V_0^*
\end{align*}
By the definition of SVD decomposition, we have $WV_0=0$, $U_1^*W=0$ and $||W||_2 \leq 1$ holds because the diagonal elements of $\Sigma_0$ is less or equal to $\tau$. So, $\hat{Z} - Y_{-i}$ is in the subgradients of $\tau ||Z||_*$ and \eqref{subgrad} holds.
\end{proof}

\section{Robust Synthetic Control Hyperparameter}
\label{appB}
\begin{figure}[]
\caption{\textbf{Eigenspectrum of West Germany Data}}
\centering
\includegraphics[width=\textwidth,height =7 cm]{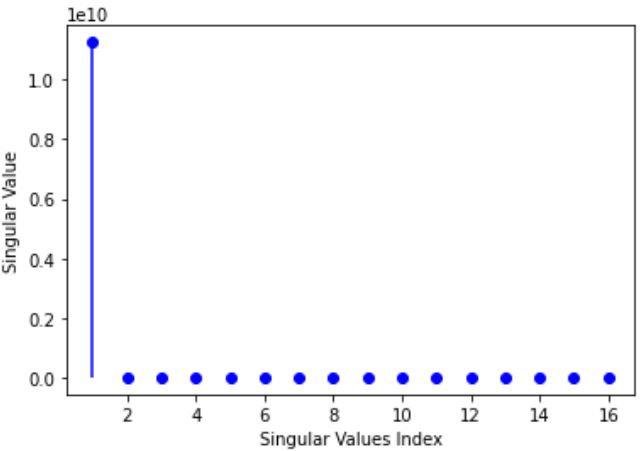}
\label{Amjad_1}
\end{figure}

\begin{figure}[]
\caption{\textbf{Accumulated Explained Variation by Singular Values}}
\centering
\includegraphics[width=\textwidth,height =7 cm]{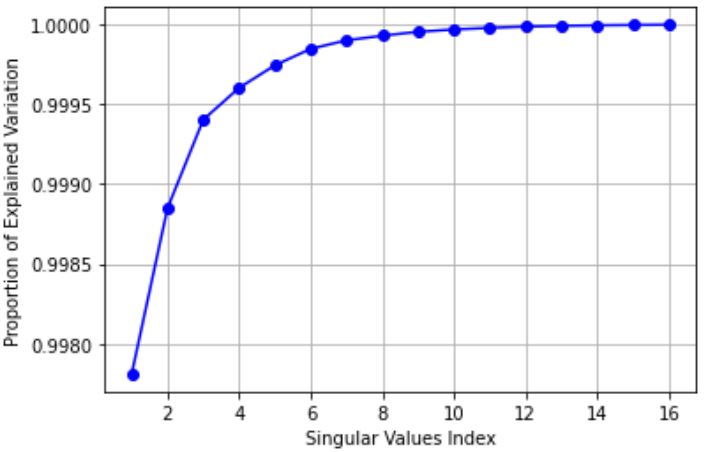}
\label{Amjad_2}
\end{figure}

As I mentioned, the robust synthetic control uses the singular values thresholding to extract the low-rank structure of the data set. Figure \ref{Amjad_1} shows the singular values of the West Germany data set. From the second singular value, these values approach zero and by considering just $2$ singular values, we can extract the low-rank structure. Figure \ref{Amjad_2} also confirms that the first $2$ singular values can explain near to $99\%$ of the variation in the data.

\end{appendices}

\newpage
\medskip
\printbibliography
\end{document}